\documentclass[preprint,12pt]{elsarticle}

\usepackage{amssymb}

\usepackage{amsmath,amssymb,amsfonts}
\usepackage{algorithmic}
\usepackage{graphicx}
\usepackage{textcomp}

\usepackage{subfigure}
\usepackage{multirow}
\usepackage{color}
\usepackage[table]{xcolor}
\usepackage{hyperref}
\usepackage{booktabs}
\usepackage{dsfont}
\usepackage{threeparttable}

\begin{document}

\begin{frontmatter}



\title{High-Fidelity 3D Gaussian Human Reconstruction via Region-Aware Initialization and Geometric Priors}


\author[label1]{Yang Liu}

\author[label1]{Zhiyong Zhang}

\address[label1]{School of Electronics and Communication Engineering, Sun Yat-sen University, Shenzhen, Guangdong, China}

\begin{abstract}
	
Real-time and high-fidelity 3D human reconstruction from RGB images is a fundamental yet challenging task, aiming to recover visually plausible digital avatars that provide core representations for downstream interactive applications such as virtual reality and gaming. 
While mainstream 3D Gaussian Splatting paradigms excel in rendering speed and static scene reconstruction, they severely struggle with the complex, time-varying non-rigid deformations of dynamic human bodies. 
Although incorporating standard parametric models provides foundational prior knowledge, existing methods inherently lack the capacity to capture high-frequency geometric details in critical regions, inevitably leading to visually disruptive artifacts such as interconnected fingers and over-smoothed facial features. 
Furthermore, decoupling dynamic attributes via standard spatial fields introduces an inherent trade-off between the preservation of fine-grained details and the exponential expenditure of GPU memory. 
To overcome the limitations of existing approaches, we propose a novel 3D Gaussian human reconstruction framework that intrinsically integrates region-aware initialization with rich geometric priors. 
Moving beyond conventional coarse initializations, our method leverages the highly expressive SMPL-X model as a generic prior to instantiate both the 3D Gaussian representations and skinning weights, thereby establishing a robust geometric foundation for high-precision reconstruction. 
Furthermore, to effectively balance rendering fidelity and computational cost, we introduce a region-aware density initialization scheme and a geometry-aware multi-scale hash encoding module, which synergistically interact to resolve localized artifacts and significantly enhance the photorealism of anisotropic Gaussian distributions across diverse poses. 
Extensive experiments on standard benchmark datasets, including PeopleSnapshot and GalaBasketball, demonstrate that our approach achieves superior reconstruction quality and fine-grained detail recovery, validating its effectiveness and enhanced robustness in dynamic scenarios with complex motions while strictly maintaining rapid rendering speeds.

\end{abstract}



\begin{keyword}


	3D Human Reconstruction\sep
	3D Gaussian Splatting\sep
	Human Avatars\sep
	Geometric Priors\sep
	Multi-Scale Hash Encoding

\end{keyword}

\end{frontmatter}



\section{Introduction} \label{sec1}

Real-time rendering and high-quality, rapid reconstruction of digital humans are widely applied in fields such as virtual reality, gaming, and sports, serving as the core of various interactive applications \cite{xie2025nsghg, nazir20263dgeomeshnet}. 
Methods based on Neural Radiance Fields (NeRF) \cite{mildenhall2021nerf} can implicitly reconstruct high-quality human avatars \cite{chen2021animatable, peng2021neural, xiu2022icon, xiu2023econ}. 
However, they are severely constrained by substantial memory and time overheads, typically requiring complex implicit representations, deformation algorithms, and volume rendering to attain high-fidelity reconstruction. 
Concurrently, several studies \cite{fridovich2022plenoxels, muller2022instant, sun2022direct, wang2022fourier} have focused on accelerating the neural rendering process. 
Nevertheless, these approaches still incur prohibitive memory costs. Furthermore, since implicit NeRFs model the entire 3D space, they inevitably introduce artifacts in synthesized novel views. 
Point sampling along each ray further exacerbates the memory and computational burden. 
Recently, the advent of 3D Gaussian Splatting \cite{kerbl20233d} has enabled rapid and high-quality 3D reconstruction \cite{li2026novel, wang2026dynamic}, marking a significant milestone for the widespread deployment of this technology across diverse domains.

Currently, there are still some challenges in applying 3D Gaussian reconstruction technology to 3D human reconstruction.
First, since the initialization of 3D Gaussians has a major impact on the quality of the optimization results, inadequate initialization may even cause the optimization process to diverge.
Moreover, the Structure-from-Motion (SfM) based initialization, which is commonly used for static 3D Gaussians, is ill-suited for the dynamic scenes inherent in human avatar reconstruction. 
Although some research works \cite{liu2024animatable} use the SMPL parametric human model \cite{loper2023smpl} as a prior for 3D Gaussian initialization, the head of the SMPL model has no facial details and the fingers of the hands are also unseparated. 
However, in human visual perception, these specific regions represent high-frequency details that draw significant attention. 
This leads to the fact that in the final 3D reconstruction results, the rendering quality of these key areas of the face and hands is often unsatisfactory, prone to severe blurring, missing details, artifacts, or geometric distortions.

Secondly, since the deformation of Gaussians from the canonical space to the posed space is initially under-determined, achieving convergence requires significantly more samples and iterations. 
Furthermore, 3D Gaussian rasterization-based rendering can only backpropagate gradients to a limited number of Gaussians during a single iteration. 
Coupled with the complex, time-varying non-rigid deformations of the human body caused by factors such as clothing dynamics, the optimization process for dynamic scenes becomes exceedingly sluggish or may even diverge. 
To address this, a hash table can be constructed to decouple the intrinsic color and displacement attributes from the Gaussian primitives, formulating them as a continuous spatial parameter field. By leveraging these spatial field constraints, the stability of dynamic scene reconstruction is substantially enhanced. 
During the hash encoding process, utilizing higher hash levels yields sharper details in critical regions like the face and hands; however, this incurs an exponential increase in GPU memory footprint. Conversely, strictly constraining memory usage inevitably leads to the over-smoothing of high-frequency details. 
Consequently, an inherent trade-off exists between the reconstruction fidelity of high-frequency details and the expenditure of computational resources in this pipeline.

To address the aforementioned issues, this paper proposes a 3D Gaussian human reconstruction approach based on region-aware initialization and geometric priors, capable of recovering high-fidelity 3D human Gaussian models from RGB image inputs. 
Instead of randomly initializing skinning weights, the proposed method leverages the SMPL-X model \cite{pavlakos2019expressive} to initialize the 3D Gaussian representations, treating the skinning weights as a generic human model prior that is independent of any specific body shape. 
Furthermore, we introduce a region-aware density initialization scheme and a geometry-aware multi-scale hash encoding module. 
These components effectively resolve the artifacts of interconnected fingers and over-smoothed facial features present in existing methods at the feature level, thereby enhancing the quality and photorealism of the anisotropic Gaussian distributions across various poses. 
In this paper, we evaluate the proposed method and compare it against baseline approaches on the widely adopted PeopleSnapshot dataset, as well as the GalaBasketball dataset, which features complex motions and dynamic shadows. 
Extensive experimental results demonstrate that our method can reconstruct high-quality 3D Gaussian human avatars while maintaining rapid rendering speeds.

The main contributions of this paper are summarized as follows:

(1) We propose a novel 3D Gaussian human reconstruction approach based on region-aware initialization and geometric priors, capable of recovering high-fidelity 3D Gaussian human avatars from RGB image inputs.

(2) We leverage the SMPL-X model as a generic human model prior to initialize both the 3D Gaussian representations and the skinning weights, thereby establishing a robust geometric foundation for high-precision reconstruction.

(3) We design a region-aware density initialization scheme and a geometry-aware multi-scale hash encoding module. These components effectively mitigate the artifacts of interconnected fingers and over-smoothed facial features, significantly enhancing the quality and photorealism of the Gaussian distributions across various poses.

(4) Extensive evaluations and comparisons against baseline methods on the PeopleSnapshot and GalaBasketball datasets demonstrate that our approach achieves high-quality 3D Gaussian human reconstruction while maintaining rapid rendering speeds.

\section{Related Work} \label{sec2}

\subsection{Mesh-based Avatar}

Recent advances in free-viewpoint video have shown that sequences of textured meshes can produce highly expressive renderings using detailed texture atlases with as few as 10k triangles \cite{collet2015high}. Building on this line of research, many studies \cite{habermann2023hdhumans,ho2023learning} have further extended such representations to controllable avatar modeling. By exploiting human body models with strong structural priors \cite{chai2022realy, li2017learning, loper2023smpl, pavlakos2019expressive}, which can be unwrapped into a shared UV space, these methods enable 2D texture atlas generation and supervision through differentiable rendering \cite{ma2021pixel,xiang2021modeling}. Such prior models provide robustness and temporal consistency under large body motions, and can be reconstructed from monocular videos or even a single image. To better capture identity-specific geometry and clothing details, CAPE \cite{ma2020learning} predicts vertex displacements via a pose-conditioned variational autoencoder. However, because the underlying body model is inherently limited in representing topological variations, some methods incorporate textured meshes as conditioning inputs for image synthesis \cite{ma2021pixel, prokudin2021smplpix}, while others adopt implicit representations to model geometry \cite{chen2021snarf, chen2023fast, ho2023learning, shen2023x}, appearance \cite{grassal2022neural, ho2023learning, shen2023x}, or material properties \cite{bharadwaj2023flare}.

\subsection{Implicit Neural Avatar}

To overcome the limitations of triangle meshes, especially in representing fine-scale structures such as hair, glasses, and loose garments, a growing body of work \cite{bai2023learning, grassal2022neural, jiang2023instantavatar, peng2021animatable, peng2021neural, xu2021h, yu2023monohuman, zielonka2023instant} constructs NeRFs in a canonical space, typically defined by the T-pose of SMPL \cite{loper2023smpl} or the neutral expression space of FLAME \cite{li2017learning}, and then performs rendering in the posed space. A central challenge in this framework is establishing correspondences by mapping points from the posed space back to the canonical space, a process that is both difficult and inherently ambiguous. To address this issue, previous methods either introduce pose-conditioned inverse linear blend skinning fields \cite{guo2023vid2avatar, peng2021animatable} or recover correspondences through optimized root-finding procedures with multiple initializations \cite{chen2021snarf, chen2023fast, jiang2023instantavatar}. Nevertheless, the substantial computational cost of volume rendering remains a major bottleneck for real-time applications.

\section{Method}

\subsection{Preliminary}

In this paper, we formulate the 3D human reconstruction problem based on Gaussian splatting as follows: given a video sequence or an image collection $I = \{I_t\}_{t=1}^T$, our goal is to reconstruct a 3D Gaussian representation for each individual via 3D Gaussian Splatting \cite{kerbl20233d}.
Each 3D Gaussian primitive $\mathcal{P}$ is parameterized as:
\begin{equation}
	\mathcal{P} = {x_0, \mathbf{R}, \mathbf{S}, \alpha_0, SH},
	\label{eq5-1}
\end{equation}
where $x_0$ denotes the geometric center of the 3D Gaussian distribution, 
$\mathbf{R}$ represents the rotation matrix, 
$\mathbf{S}$ is the scaling matrix defining the spatial scale along three dimensions, 
$\alpha_0$ denotes the base opacity, 
and $SH$ represents a set of spherical harmonics coefficients used to model the view-dependent color distribution, following standard practices \cite{fridovich2022plenoxels}.
The opacity at a spatial position $x$ near the 3D Gaussian $\mathcal{P}$ is evaluated as:
\begin{equation}
	\alpha(x)=\alpha_0e^{-\frac{1}{2}(x-x_0)^T\Sigma^{-1}(x-x_0)},
	\label{eq5-2}
\end{equation}
where the covariance matrix $\Sigma$ is factorized into the rotation matrix $\mathbf{R}$ and the scaling matrix $\mathbf{S}$:
\begin{equation}
	\Sigma=\mathbf{R}\mathbf{S}\mathbf{S}^T\mathbf{R}^T.
	\label{eq5-3}
\end{equation}
The 3D Gaussians are projected onto the 2D image plane utilizing the Elliptical Weighted Average (EWA) volume splatting algorithm \cite{zwicker2001ewa}:
\begin{equation}
	\Sigma^{^{\prime}}=\mathbf{P}\mathbf{W}\Sigma \mathbf{W}^T\mathbf{P}^T,
	\label{eq5-4}
\end{equation}
where $\mathbf{W}$ denotes the viewing transformation matrix, 
and $\mathbf{P}$ is the Jacobian matrix representing the affine approximation of the projective transformation.
The projected Gaussians are subsequently sorted by depth and rasterized over the overlapping pixels.
For a given pixel, the rasterizer accumulates a depth-sorted list of colors $c_i$ and opacities $\alpha_i$.
The final pixel color $C$ is computed via alpha blending of the $N$ ordered points covering that pixel:
\begin{equation}
	C=\sum_{i\in N}c_i\alpha_i\prod_{j=1}^{i-1}(1-\alpha_j).
	\label{eq5-5}
\end{equation}
Building upon this 3D Gaussian representation, when provided with a target pose as the driving condition, we apply skeletal transformations to spatially deform the canonical Gaussian primitives. Subsequently, these deformed Gaussians are projected onto the 2D imaging plane via 3D Gaussian splatting to synthesize the final rendered image.

\subsection{Overall Architecture}

\begin{figure}[htbp]
	\centering
	\includegraphics[width=1\textwidth]{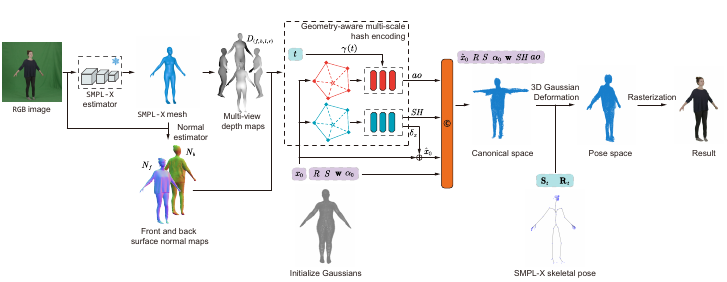}
	\caption{An overview of our proposed framework.} 
	\label{fig5-1}
\end{figure}

As illustrated in Figure \ref{fig5-1}, the human avatar reconstruction framework proposed in this paper aims to recover high-fidelity, animatable 3D human models from RGB images. 
Our framework consists of four core stages: human geometric prior extraction, geometry-aware multi-scale hash encoding, canonical space Gaussian modeling, and pose-guided deformation and rendering. 
By integrating the robust priors of the SMPL-X parametric human model with the explicit representation capabilities of 3D Gaussian splatting, our method effectively models complex human geometry and appearance.

Specifically, the RGB image is first fed into a pre-trained, frozen SMPL-X estimator to predict the underlying SMPL-X mesh of the human body. 
Subsequently, multi-view depth rendering is applied to this mesh to extract depth maps from four canonical viewpoints (front, back, left, and right), denoted as $D_{\{f,b,l,r\}}$. 
Simultaneously, a normal estimation network \cite{cao2022bilateral} is employed to infer the front and back surface normal maps from the input image, denoted as $N_f, N_b$, thereby providing crucial spatial geometric constraints for the subsequent networks. 
To efficiently capture intricate human details, the extracted depth and normal geometric features are forwarded to the geometry-aware multi-scale hash encoding module. 
This module takes the initial position $x_0$ and the positionally encoded time $\gamma(t)$ as inputs. 
It queries and maps features utilizing a multi-scale hash grid, and ultimately employs a Multi-Layer Perceptron (MLP) to predict the residual attributes that compensate the 3D Gaussian representation, encompassing the position offset $\delta_x$, spherical harmonics features $SH$, and the ambient occlusion parameter $ao$. 
The canonical space Gaussian modeling stage is responsible for constructing a comprehensive Gaussian representation within a unified Canonical Space. 
The initialized 3D Gaussians are sampled from the SMPL-X template surface, possessing a foundational parameter set: initial position $x_0$, rotation $\mathbf{R}$, scaling $\mathbf{S}$, non-rigid deformation weights $\mathbf{w}$, and opacity $\alpha_0$. 
The network-predicted position offset $\delta_x$ is then added to the initial position $x_0$ to yield the fine-tuned Gaussian center $\hat{x}_0$. 
Following this, the corrected position $\hat{x}_0$, the initial geometric parameters ($\mathbf{R}, \mathbf{S}, \alpha_0, \mathbf{w}$), and the predicted appearance parameters ($SH, ao$) are concatenated along the channel dimension to constitute the complete canonical-space Gaussian model. 
Finally, the framework maps the canonical model into the posed space, and projects these transformed 3D Gaussians onto the 2D image plane via splatting to synthesize high-quality rendered results.

Our proposed representation accurately fits dynamic scenes involving moving individuals, thus obviating the need for auxiliary regularization losses. The overall loss function is formulated as a direct combination of the $L_1$ loss and a Distance-Structural Similarity Index Measure (D-SSIM) term, $L_{D-SSIM}$:
\begin{equation}
	L = (1 - \lambda)L_1 + \lambda L_{D-SSIM},
	\label{eq5-11}
\end{equation}
where $\lambda$ is empirically set to 0.2, following the optimization configurations of previous works \cite{kerbl20233d, liu2024animatable}.

\subsection{Region-Aware Density Initialization} \label{sec3.3}	

To facilitate 3D Gaussian deformation based on the Linear Blend Skinning (LBS) algorithm, during our canonical space modeling process, we construct a set of skinned 3D Gaussians and bind them to the corresponding skeleton $J$. 
The formulated skinned 3D Gaussian $P_{skin}$ is reconstructed from the static 3D Gaussian primitive $P$ (Eq. \eqref{eq5-1}), parameterized as:
\begin{equation}
	P_{skin} = {x_0, \mathbf{R}, \mathbf{S}, \alpha_0, SH, \mathbf{w}, \delta_x},
	\label{eq5-12}
\end{equation}
where $w$ denotes the skinning weights and $\delta_x$ represents the vertex displacement.
Unlike previous works \cite{liu2024animatable} that utilize the SMPL model \cite{loper2023smpl} to extract the initial 3D Gaussian point cloud, we propose a region-aware density initialization scheme based on the SMPL-X model \cite{pavlakos2019expressive}. 
This aims to ensure sufficient points for fitting high-frequency details in the facial and hand regions, thereby mitigating the artifacts of interconnected fingers or over-smoothed facial features during rendering. 
Leveraging the finer-grained parametric representation of SMPL-X, we increase the initial density of Gaussian primitives in these areas during surface sampling, allocating greater representational capacity accordingly.

To address the geometric tearing and texture seams at the junctions between the torso and the extremities (hands/face) caused by varying hash resolutions, we introduce a smooth soft-mask transition mechanism based on mesh geodesic distance.
Let $V_{high}$ denote a predefined set of core vertices representing the face and hands. For any query point $x$ on the model surface, we compute its shortest geodesic distance to $V_{high}$, denoted as $d(x, V_{high})$. 
Based on this, we formulate a continuous region-aware weight $\tau(x) \in [0,1]$ as:
\begin{equation}
	\tau(x) = \exp \left( - \frac{d(x, V_{high})^2}{2\sigma^2} \right),
	\label{eq5-31}
\end{equation}
where $\sigma$ is a hyperparameter controlling the width of the transition band.
During hash feature querying within this transition band, our method concurrently retrieves the low-resolution feature $F_{l}$ and the high-resolution feature $F_{h}$, blending them via linear interpolation guided by $\tau(x)$:
\begin{equation}
	F_{f}(x) = \tau(x) F_{h}(x) + (1 - \tau(x)) F_{l}(x).
	\label{eq5-32}
\end{equation}
Through this continuous feature interpolation, our approach guarantees $C^0$ continuity of the parameter field at semantic boundaries, effectively eliminating high-frequency seam artifacts in the rendered results.

Generally, upsampling the input model vertices to approximately 100,000 is sufficient to achieve high-quality reconstruction.
Based on the SMPL-X mesh, we randomly sample $K$ additional points within the local neighborhood of each vertex, directly duplicating their skinning parameters.
To prevent prominent Gaussian patches caused by abrupt local density mutations, we eschew a hard assignment of $K$ during initialization. 
Instead, we formulate the sampling count $K$ as a continuous spatial function:
\begin{equation}
	K(x) =  K_{b} + (K_{m} - K_{b}) \cdot \tau(x),
	\label{eq5-33}
\end{equation}
where $K_{b}=15$ for the torso region, and $K_{m}=20$ for the facial and hand regions.
Consequently, the density of Gaussian primitives exhibits a natural gradient increase from the arms to the wrists and palms. 
This allows the network to allocate hash table capacity more smoothly, ultimately yielding a model with approximately 170,000 points.

Thus, during canonical space modeling, the skinned 3D Gaussian model encapsulates a set of positions, skinning weights $w$, and a skeleton $J$ corresponding to the input pose and facial/hand parameters. 
During optimization, the skinning weights $w$ are kept fixed, while the vertex displacement $\delta_x$ and the skeleton $J$ are jointly optimized. 
This strategy ensures the accurate capture of human shapes and motions. Prior to pose-guided deformation, an offset is applied to the canonical Gaussian center $x_0$:
\begin{equation}
	x'_0 = x_0 + \delta_x.
	\label{eq5-13}
\end{equation}

\subsection{Shape and Appearance Hash Encoding} \label{sec3.4}

To capture intricate details in the facial and hand regions while minimizing GPU memory consumption, we propose a geometry-aware multi-scale hash encoding for shape and appearance. 
This approach achieves parameter decoupling via a multi-scale strategy: the torso and garments utilize low-frequency hash representations, whereas high-frequency hash queries are exclusively activated for the face and hands.
During inference, if a given point belongs to the facial or hand regions, higher-resolution hash levels are queried, enabling the reconstruction of higher-frequency details. 
Furthermore, auxiliary depth and normal features are incorporated during the computation of hash features, substantially bolstering the network's global contextual awareness and collision perception capabilities.

Specifically, we sample the spherical harmonic (SH) coefficients $SH$ from a continuous parameter field for each vertex, effectively influencing all adjacent Gaussian primitives within a single optimization step.
Additionally, to prevent optimization divergence in dynamic scenes caused by unconstrained per-vertex displacements, we also model a continuous parameter field for the vertex displacements.
To account for complex motions in environments with dynamic lighting changes, we enable dynamic ambient occlusion \cite{liu2024animatable}. 
The spatial position $x_0$ of the Gaussian primitive, combined with the current timestamp $t$, is fed into the hash grid and an MLP to query the instantaneous shading intensity.
Consequently, the Gaussian primitives are decoupled from explicit color attributes; instead, their SH coefficients and displacements are actively queried from the hash table based on their spatial coordinates $(x,y,z)$.
The remaining parameters defined in Eq. \eqref{eq5-12} are explicitly stored within each primitive to preserve the flexibility of 3D Gaussians in fitting arbitrary shapes. 
Ultimately, the parameter field can be formulated as:
\begin{equation}
	x_0 \mapsto SH, \delta_x.
	\label{eq5-14}
\end{equation}

Since our 3D Gaussian representation is initialized from a canonical human mesh model, the centers of the Gaussian primitives are uniformly distributed in proximity to the human body surface.
Because spatially adjacent Gaussians query the same or neighboring entries in the hash grid, updating the parameters at a fixed sampled location near the surface implicitly corrects the appearance of all adjacent Gaussians within that local region simultaneously.
Building upon this premise, we compress the underlying multi-resolution hash table \cite{muller2022instant}, significantly mitigating both computational overhead and storage consumption by explicitly modeling the parameter field.
As the hash table intrinsically stores exceedingly rich spatial details, processing sparse volumetric data requires only a lightweight MLP to decode the final results, thereby drastically reducing the computational burden.

To explicitly incorporate the extracted geometric priors into parameter field learning, for an arbitrary point $x_0$ in canonical space, we first project it onto four depth-map planes corresponding to the front, back, left, and right views, as well as two normal-map planes corresponding to the front and back views, using orthographic projections aligned with the SMPL-X mesh.

Let $v\in\{f,b,l,r\}$ denote a depth view, and let $\pi_v(x_0)$ represent the projection of point $x_0$ under view $v$.
If the projected location falls within the valid human-body region, the corresponding depth value is sampled from the depth map $D_v$ via bilinear interpolation, yielding the multi-view depth feature:
\begin{equation}
	\phi_d(x_0)=\big[d_f(x_0),d_b(x_0),d_l(x_0),d_r(x_0)\big],
	\label{eq5-34}
\end{equation}
where $d_v(x_0)$ denotes the local depth sample of point $x_0$ under view $v$.
Similarly, for the front and back normal maps, the normal feature is obtained through the same projection and bilinear sampling procedure:
\begin{equation}
	\phi_n(x_0)=\big[n_f(x_0),n_b(x_0)\big].
	\label{eq5-35}
\end{equation}
Here, $n_f(x_0)$ and $n_b(x_0)$ denote the local normal samples of point $x_0$ from the front and back normal maps, respectively.
If the projected location lies outside the valid human region, the corresponding feature is padded with a zero vector to ensure a consistent network input dimensionality.

After constructing these geometric priors, we feed the multi-scale hash-grid features, depth features, normal features, and temporal encoding jointly into a multilayer perceptron.
For a point $x_0$ in canonical space, let $\mathcal{H}(x_0)$ denote the feature queried from the multi-scale hash grid at that location, and let $\gamma(t)$ denote the temporal encoding. The joint input of the geometry-aware hash encoding is then formulated as:
\begin{equation}
	z(x_0,t)=\big[\mathcal{H}(x_0),\phi_d(x_0),\phi_n(x_0),\gamma(t)\big].
	\label{eq5-36}
\end{equation}
A lightweight multilayer perceptron then predicts the residual attributes associated with the current point:
\begin{equation}
	(\delta_x,SH)=\mathrm{MLP}\big(z(x_0,t)\big).
	\label{eq5-37}
\end{equation}

Specifically, the multi-view depth features provide positional constraints on the Gaussian center with respect to the global geometric envelope of the human body, enabling the network to determine whether the current point deviates from the body surface and thereby stabilizing the learning of the displacement $\delta_x$. The front and back normal features, in contrast, encode local surface orientation cues, improving the network's ability to resolve clothing wrinkles, limb boundaries, self-occluded regions, and thin structures.
With these geometry-conditioned constraints, the network no longer relies solely on local coordinates and hash indices when learning positional offsets and appearance attributes; instead, it explicitly leverages priors derived from the overall human surface structure. This design effectively reduces floating points, geometric tearing, and local artifacts, while improving fine-detail reconstruction in challenging regions.

\subsection{3D Gaussian Deformation} \label{sec3.5}

In this paper, we employ a skinning weight field \cite{pavlakos2019expressive} to model joint articulation. 
Each frame in the video sequence is formulated as a posed space indexed by timestamp $t$, and the canonical points are deformed into the posed space via Linear Blend Skinning (LBS). 
The skinning weight field is defined as:
\begin{equation}
	\mathbf{w}(x_c)=\{w_1,...w_{n_b}\}, 
	\label{eq5-15}
\end{equation}
where $n_b$ denotes the total number of skeletal joints, and $x_c$ represents a sampled point in the canonical space.
Consequently, the target bone transformations $\mathbf{B}_{t}=\{B_{1}^{t},...,B_{n_{b}}^{t}\}$ at frame $t$ can be derived from the input pose and the corresponding skeleton configuration:
\begin{equation}
	\mathbf{S}_t,\mathbf{J},\mathbf{T}_t\mapsto\mathbf{B}_t,
	\label{eq5-16}
\end{equation}
where $\mathbf{S}_t = {\omega_{1}^{t}, ..., \omega_{n_{b}}^{t}}$ denotes the Euler angles for the rotation of each joint at frame $t$ (with $\omega_{1}^{t}$ corresponding to the global rotation in the world coordinate system, and the remainder being local joint rotations); $\mathbf{T}_t$ represents the global translation at frame $t$; and $\mathbf{J}=\{J_{1},...,J_{n_{b}}\}$ indicates the local coordinates of each joint within the canonical space.
Subsequently, the deformed position $x_t$ in the posed space at frame $t$ can be obtained by applying LBS to the canonical point $x_c$:
\begin{equation}
	x_t=\sum_{i=1}^{n_b}w_iB_i^tx_c.
	\label{eq5-17}
\end{equation}

\begin{figure}[htbp]
		\centering
		\includegraphics[width=0.6\textwidth]{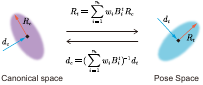}
		\caption{The details of 3D Gaussian deformation.} 
		\label{fig5-2}
	\end{figure}

To capture intricate details, we leverage the SMPL-X model, which incorporates not only body shape and pose parameters but also expression parameters, jaw articulation, and hand joint rotations. 
This effectively expands the number of driving joints from 24 to 54.
Maintaining a static orientation across varying poses would cause the 3D Gaussians to erroneously degenerate into isotropic spheres. 
Thus, it is imperative to dynamically update their orientations to preserve consistent anisotropic Gaussian distributions under diverse poses.
As illustrated in Figure \ref{fig5-2}, LBS is identically applied to the rotation matrix $\mathbf{R}$ of the 3D Gaussian primitive:
\begin{equation}
	\mathbf{R}_t=\sum_{i=1}^{n_b}w_iB_i^t\mathbf{R}_c,
	\label{eq5-18}
\end{equation}
where $\mathbf{R}_c$ is the rotation in the canonical space, and $\mathbf{R}_t$ is the posed rotation at frame $t$.
Furthermore, to compute view-dependent colors via spherical harmonics, the viewing direction must be inversely warped back to the canonical space to maintain anisotropic color consistency.
Therefore, the inverse LBS transformation is applied to the viewing direction:
\begin{equation}
	d_c=(\sum_{i=1}^{n_b}w_iB_i^t)^{-1}d_t,
	\label{eq5-19}
\end{equation}
where $d_c$ denotes the direction in the canonical space, and $d_t$ represents the viewing direction in the posed space at frame $t$.
By extending the official 3D Gaussian rasterizer \cite{kerbl20233d} and explicitly deriving the analytical gradients for all parameterized variables, we seamlessly implement these transformations within the rendering pipeline. Consequently, the computational overhead introduced by the pose-guided 3D Gaussian deformation is practically negligible.

\subsection{Ambient Occlusion Rendering} \label{sec3.6}

To account for temporal dynamics during the rendering of dynamic scenes, we incorporate a time-varying ambient occlusion module by explicitly modeling an occlusion factor $ao \in [0,1]$ for each primitive:
\begin{equation}
	P_{skin}^{^{\prime}}=\{x_0,\mathbf{R},\mathbf{S},\alpha_0,SH,\mathbf{w},\delta_x,ao\}.
	\label{eq5-20}
\end{equation}
Once the ambient occlusion factor is computed for each individual Gaussian, the final view-dependent color $c$ is evaluated as:
\begin{equation}
	c=ao\cdot\mathbf{Y}(SH,d_c),
	\label{eq5-21}
\end{equation}
where $\mathbf{Y}(\cdot)$ denotes the evaluation of the spherical harmonics (SH) given the canonical viewing direction $d_c$.
Similarly, we adopt multi-resolution hash encoding \cite{muller2022instant} for the ambient occlusion term $ao$. To effectively capture time-varying shadowing effects, the positional encoding of the timestamp $t$ is concatenated as an auxiliary input to the MLP of the hash encoder. Consequently, the parameter field governing the ambient occlusion is formulated as:
\begin{equation}
	x_0,\gamma(t)\mapsto ao,
	\label{eq5-22}
\end{equation}
where $\gamma(\cdot)$ denotes the standard high-frequency positional encoding function introduced in NeRF \cite{mildenhall2021nerf}.

\section{Experiments}

\subsection{Experimental Setup}
\subsubsection{Datasets and Evaluation Metrics}

We employ the PeopleSnapshot dataset \cite{alldieck2018detailed} to evaluate our model's performance in single-person, monocular scenarios. 
Furthermore, the GalaBasketball dataset \cite{liu2024animatable} is utilized to assess its capabilities in multi-person dynamic scenes and novel view synthesis for single-person settings.
The PeopleSnapshot dataset comprises extensive visual data of individuals wearing casual clothing, effectively overcoming the limitations of traditional laboratory datasets that rely exclusively on dedicated motion-capture suits.
Within these monocular sequences, subjects of diverse genders, body shapes, and clothing types perform 360-degree rotations and simple motions in front of the camera.
Each video frame is provided with high-quality ground-truth reconstructions and corresponding pose parameters.
The GalaBasketball dataset is synthesized using several player models with varying shapes and appearances. 
It encompasses four single-person scenes and three multi-person scenes featuring complex articulations and dynamic shadows. It provides six uniformly distributed camera viewpoints for training and one reserved for testing.

To quantitatively evaluate the rendering quality, we adopt three standard metrics: Peak Signal-to-Noise Ratio (PSNR), Structural Similarity Index Measure (SSIM) \cite{wang2004image}, and Learned Perceptual Image Patch Similarity (LPIPS) \cite{zhang2018unreasonable}.
PSNR is computed based on the Mean Square Error (MSE) between the synthesized and ground-truth pixels with respect to the maximum possible pixel value; higher values (in dB) indicate superior image fidelity.
SSIM evaluates the similarity based on the high sensitivity of the human visual system to local structural variations, comprehensively measuring luminance, contrast, and structure. Scores range from 0 to 1 (where 1 indicates identical images), with higher values reflecting better structural preservation.
LPIPS feeds the rendered and ground-truth images into a pre-trained Convolutional Neural Network (CNN) to extract and compute their distance in the deep feature space. A score of 0 implies perceptual equivalence, whereas higher values indicate greater perceptual discrepancies.

\subsubsection{Baselines}

We compare our method against state-of-the-art baselines, including the NeRF-based Anim-NeRF \cite{chen2021animatable}, as well as InstantAvatar \cite{jiang2023instantavatar} and Animatable 3D Gaussian \cite{liu2024animatable}, which are based on 3D Gaussian representations.

\subsubsection{Implementation Details}

For a fair comparison, we adhere to the experimental protocols established in prior works \cite{jiang2023instantavatar,liu2024animatable}. The input images are downsampled to a resolution of $540 \times 540$, and the models are trained solely on estimated body parameters without utilizing any ground-truth counterparts. All evaluations are conducted on a single NVIDIA RTX 4090 GPU.
Given that the PeopleSnapshot dataset lacks temporal metadata and drastic shadow variations, we disable the time-varying dynamic ambient occlusion (AO) module for these specific sequences.
Conversely, on the GalaBasketball dataset, a fixed AO factor is applied during the initial training phase. This strategy forces the model to learn time-independent spherical harmonic (SH) coefficients first. The active optimization of the dynamic AO module commences only after the rendered colors have fully stabilized.

\subsection{Results and Analysis}

\subsubsection{Evaluation on the PeopleSnapshot Dataset}

To objectively assess our method's capabilities in monocular human reconstruction, we conduct quantitative evaluations on four sequences from the PeopleSnapshot dataset. The detailed comparative results are summarized in Table \ref{tab5-1}.
Under a training time budget of approximately 30 seconds, our approach consistently achieves state-of-the-art performance across all evaluation metrics.
First, compared to the traditional implicit NeRF-based method, Anim-NeRF, which requires approximately 5 minutes to train, our method dramatically surpasses its rendering quality while achieving a nearly 10-fold acceleration in training speed. For instance, on the \textit{male-3-casual} sequence, our model yields a substantial PSNR improvement of 7.37 dB. This forcefully demonstrates the immense efficiency advantages inherent to the explicit representations of 3D Gaussians.
Furthermore, compared to the recent Animatable 3D Gaussian method—which shares the underlying 3DGS architecture—our approach significantly boosts reconstruction accuracy without compromising rendering speed.
Specifically, on the \textit{male-4-casual} sequence, our method attains a PSNR of 28.88 dB (an absolute gain of 2.72 dB over Animatable 3D Gaussian) and reduces the LPIPS error to 0.0401.
This profound enhancement in perceptual quality is primarily attributed to our proposed SMPL-X-based region-aware density initialization and the geometry-aware multi-scale hash encoding strategy. 
By explicitly allocating greater representational capacity to the intricate facial and hand regions, and leveraging auxiliary depth and normal features to guide local detail learning, our model fits the high-frequency textures and complex surface geometries with unprecedented precision.

To further validate the convergence efficiency and robustness of our model under extreme temporal constraints, we evaluate the reconstruction metrics of all baselines after merely 5 seconds of training, as presented in the lower half of Table \ref{tab5-1}.
Under this drastically compressed optimization window, InstantAvatar suffers from severe metric degradation and heavily distorted renderings. Conversely, our method comprehensively outperforms Animatable 3D Gaussian across all four test sequences.
Remarkably, on the \textit{female-3-casual} sequence, our method achieves a PSNR of 21.86 dB after just 5 seconds, nearly matching the 30-second training performance of Animatable 3D Gaussian, while strictly maintaining superior SSIM and LPIPS scores.
These results firmly corroborate that integrating robust human geometric priors endows the network with explicit global orientation awareness and spatial constraints. Consequently, this significantly shrinks the search space of the MLP, enabling the continuous parameter field to converge rapidly to a plausible local optimum within an extremely brief timeframe.

	\begin{table}[htbp]
	\centering
	\caption{Performance comparison of the proposed method on the PeopleSnapshot dataset.}
	\resizebox{1\textwidth}{!}{
		\begin{tabular}{lcccccccccccc}
			\toprule[1pt]
			& \multicolumn{3}{c}{male-3-casual}                   & \multicolumn{3}{c}{male-4-casual}                   & \multicolumn{3}{c}{female-3-casual}                 & \multicolumn{3}{c}{female-4-casual}                 \\ \cmidrule[0.75pt]{2-13} 
			& PSNR$\uparrow$(dB) & SSIM$\uparrow$ & LPIPS$\downarrow$ & PSNR$\uparrow$ (dB) & SSIM$\uparrow$ & LPIPS$\downarrow$ & PSNR$\uparrow$(dB) & SSIM$\uparrow$ & LPIPS$\downarrow$ & PSNR$\uparrow$(dB) & SSIM$\uparrow$ & LPIPS$\downarrow$ \\ \cmidrule[0.75pt]{1-13} 
			Anim-NeRF \cite{chen2021animatable} ($\sim$5 minutes)               & 23.17          & 0.9266         & 0.0784            & 22.30           & 0.9235         & 0.0911            & 22.37          & 0.9311         & 0.0784            & 23.18          & 0.9292         & 0.0687            \\
			InstantAvatar \cite{jiang2023instantavatar} ($\sim$30 seconds)          & 26.56          & 0.9301         & 0.1190             & 26.10           & 0.9289         & 0.1397            & 22.37          & 0.8427         & 0.2687            & 26.32          & 0.9281         & 0.1333            \\
			Animatable 3D Gaussian \cite{liu2024animatable} ($\sim$30 seconds) & 29.06          & 0.9704         & 0.0264            & 26.16          & 0.9554         & 0.0491            & 24.59          & 0.9535         & 0.0399            & 27.26          & 0.9634         & 0.0281            \\
			\cellcolor[rgb]{.851,.851,.851}Ours    ($\sim$30 seconds)                & \cellcolor[rgb]{.851,.851,.851}\textbf{30.54}          & \cellcolor[rgb]{.851,.851,.851}\textbf{0.9719}         & \cellcolor[rgb]{.851,.851,.851}\textbf{0.0232}            & \cellcolor[rgb]{.851,.851,.851}\textbf{28.88}          & \cellcolor[rgb]{.851,.851,.851}\textbf{0.9608}         & \cellcolor[rgb]{.851,.851,.851}\textbf{0.0401}            & \cellcolor[rgb]{.851,.851,.851}\textbf{25.89}          & \cellcolor[rgb]{.851,.851,.851}\textbf{0.9561}         & \cellcolor[rgb]{.851,.851,.851}\textbf{0.0315}            & \cellcolor[rgb]{.851,.851,.851}\textbf{27.33}          & \cellcolor[rgb]{.851,.851,.851}\textbf{0.9722}         & \cellcolor[rgb]{.851,.851,.851}\textbf{0.0217}            \\ \midrule[0.75pt]
			InstantAvatar \cite{jiang2023instantavatar} ($\sim$5 seconds)           & 17.30           & 0.7980          & 0.3473            & 16.44          & 0.7960          & 0.3508            & 17.22          & 0.8185         & 0.3262            & 14.91          & 0.7606         & 0.3878            \\
			Animatable 3D Gaussian \cite{liu2024animatable} ($\sim$5 seconds)  & 22.84          & 0.9395         & 0.0632            & 18.88          & 0.9103         & 0.1175            & 20.51          & 0.9301         & 0.0764            & 20.63          & 0.9246         & 0.0737            \\
			\cellcolor[rgb]{.851,.851,.851}Ours    ($\sim$5 seconds)                 & \cellcolor[rgb]{.851,.851,.851}\textbf{23.06}          & \cellcolor[rgb]{.851,.851,.851}\textbf{0.9552}         & \cellcolor[rgb]{.851,.851,.851}\textbf{0.0501}            & \cellcolor[rgb]{.851,.851,.851}\textbf{20.10}           & \cellcolor[rgb]{.851,.851,.851}\textbf{0.9252}         & \cellcolor[rgb]{.851,.851,.851}\textbf{0.0822}            & \cellcolor[rgb]{.851,.851,.851}\textbf{21.86}          & \cellcolor[rgb]{.851,.851,.851}\textbf{0.9477}         & \cellcolor[rgb]{.851,.851,.851}\textbf{0.0651}            & \cellcolor[rgb]{.851,.851,.851}\textbf{21.85}          & \cellcolor[rgb]{.851,.851,.851}\textbf{0.9481}         & \cellcolor[rgb]{.851,.851,.851}\textbf{0.0712}            \\ \bottomrule[1pt]
		\end{tabular}
	}
	\label{tab5-1}
	\end{table}

\begin{figure}[htbp]
		\centering
		\includegraphics[width=1\textwidth]{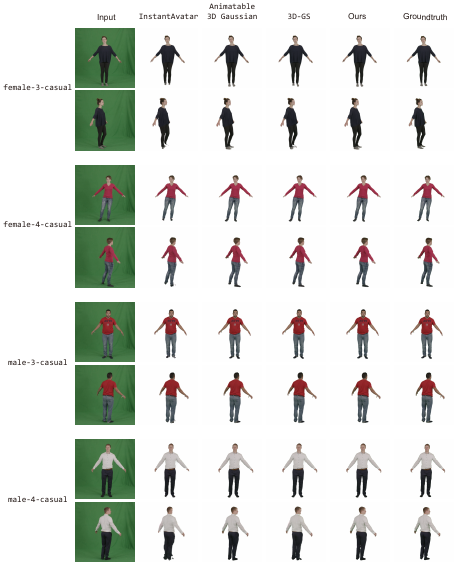}
		\caption{Qualitative comparisons on the PeopleSnapshot dataset.} 
		\label{fig5-3}
	\end{figure}

To provide a more intuitive assessment of the visual fidelity achieved by our proposed reconstruction framework, Figure \ref{fig5-3} presents the qualitative comparisons of our method against InstantAvatar, 3D-GS, and Animatable 3D Gaussian on the PeopleSnapshot dataset.
As observed from the visual results, when tackling highly challenging high-frequency details, InstantAvatar, 3D-GS, and Animatable 3D Gaussian still suffer from noticeable detail degradation, severe edge blurring, and limb-sticking artifacts.
These qualitative comparisons align seamlessly with the quantitative metrics reported in Table \ref{tab5-1}. This convincingly demonstrates that in the monocular human reconstruction task, our method not only guarantees remarkable reconstruction efficiency but also synthesizes high-fidelity human avatars enriched with intricate geometric and textural details.

\subsubsection{Evaluation on the GalaBasketball Dataset}

To assess our method's reconstruction capabilities in challenging scenarios characterized by complex limb articulations and dynamic shading variations, we conduct further evaluations on the GalaBasketball dataset, with quantitative results summarized in Table \ref{tab5-2}.
Under the baseline configuration without dynamic ambient occlusion (w/o $ao$), our approach achieves remarkable convergence within approximately 50 seconds, significantly outperforming both the similarly configured Animatable 3D Gaussian and the computationally demanding InstantAvatar.
Particularly, on the highly dynamic \textit{dribble} sequence, our method attains a PSNR of 38.82 dB, demonstrating an absolute margin of 2.29 dB over Animatable 3D Gaussian (w/o $ao$).
This substantial improvement unequivocally validates that our incorporated depth and normal geometric priors explicitly enhance the network's awareness of limb self-occlusions and collisions under intricate poses. Consequently, even without active illumination compensation, the spatial deformations of the 3D Gaussians remain highly accurate and robust.
Upon the integration of the temporal-encoded ambient occlusion and hash-based spherical harmonic (hash-SH) features, our framework consistently achieves a PSNR exceeding 40 dB across all four testing sequences. 
Furthermore, the SSIM escalates to 0.9962, while the LPIPS diminishes to 0.0023.
Compared to Animatable 3D Gaussian (hash-SH), our method exhibits distinctly superior detail recovery capabilities within a mere 70 seconds of training.
These compelling results indicate that by jointly optimizing the time-varying $ao$ factor alongside high-frequency hash features, our model can precisely capture the intricate dynamic shadows induced by basketball motions, thereby dramatically elevating the physical realism of the synthesized avatars.

	\begin{table}[htbp]
	\centering
	\caption{Performance comparison of the proposed method on the GalaBasketball dataset.}
	\resizebox{1\textwidth}{!}{
		\begin{tabular}{lcccccccccccc}
			\toprule[1pt]
			& \multicolumn{3}{c}{idle}                            & \multicolumn{3}{c}{dribble}                         & \multicolumn{3}{c}{shot}                            & \multicolumn{3}{c}{turn}                            \\ \cmidrule[0.75pt]{2-13} 
			& PSNR$\uparrow$(dB) & SSIM$\uparrow$ & LPIPS$\downarrow$ & PSNR$\uparrow$(dB) & SSIM$\uparrow$ & LPIPS$\downarrow$ & PSNR$\uparrow$(dB) & SSIM$\uparrow$ & LPIPS$\downarrow$ & PSNR$\uparrow$(dB) & SSIM$\uparrow$ & LPIPS$\downarrow$ \\ \midrule[0.75pt]
			InstantAvatar \cite{jiang2023instantavatar} ($\sim$5 minutes)                      & 29.86          & 0.9607         & 0.0575            & 27.25          & 0.9435         & 0.0903            & 29.22          & 0.9461         & 0.0709            & 32.20           & 0.9705         & 0.0371            \\
			Animatable 3D Gaussian \cite{liu2024animatable}: w/o $ao$ ($\sim$50   seconds)  & 37.38          & 0.9941         & 0.0042            & 36.53          & 0.9909         & 0.0059            & 37.07          & 0.9908         & 0.0067            & 36.77          & 0.9927         & 0.0062            \\
			Animatable 3D Gaussian \cite{liu2024animatable}: hash-SH ($\sim$70   seconds) & 39.85          & 0.9957         & 0.0032            & 38.52          & 0.9934         & 0.0043            & 39.11          & 0.9933         & 0.0051            & 39.25          & 0.9951         & 0.0041            \\
			\cellcolor[rgb]{.851,.851,.851}Ours: w/o $ao$ ($\sim$50 seconds)                      & \cellcolor[rgb]{.851,.851,.851}38.51          & \cellcolor[rgb]{.851,.851,.851}0.9955         & \cellcolor[rgb]{.851,.851,.851}0.0031            & \cellcolor[rgb]{.851,.851,.851}38.82          & \cellcolor[rgb]{.851,.851,.851}0.9921         & \cellcolor[rgb]{.851,.851,.851}0.0042            & \cellcolor[rgb]{.851,.851,.851}38.72          & \cellcolor[rgb]{.851,.851,.851}0.9917         & \cellcolor[rgb]{.851,.851,.851}0.0057            & \cellcolor[rgb]{.851,.851,.851}39.25          & \cellcolor[rgb]{.851,.851,.851}0.9937         & \cellcolor[rgb]{.851,.851,.851}0.0057            \\
			\cellcolor[rgb]{.851,.851,.851}Ours: hash-SH ($\sim$70 seconds)                     & \cellcolor[rgb]{.851,.851,.851}\textbf{41.26}          & \cellcolor[rgb]{.851,.851,.851}\textbf{0.9962}         & \cellcolor[rgb]{.851,.851,.851}\textbf{0.0023}            & \cellcolor[rgb]{.851,.851,.851}\textbf{40.50}           & \cellcolor[rgb]{.851,.851,.851}\textbf{0.9967}         & \cellcolor[rgb]{.851,.851,.851}\textbf{0.0038}            & \cellcolor[rgb]{.851,.851,.851}\textbf{41.07}          & \cellcolor[rgb]{.851,.851,.851}\textbf{0.9964}         & \cellcolor[rgb]{.851,.851,.851}\textbf{0.0046}            & \cellcolor[rgb]{.851,.851,.851}\textbf{41.01}          & \cellcolor[rgb]{.851,.851,.851}\textbf{0.9962}         & \cellcolor[rgb]{.851,.851,.851}\textbf{0.0035}            \\ \bottomrule[1pt]
		\end{tabular}
	}
	\label{tab5-2}
	\end{table}

\begin{figure}[htbp]
		\centering
		\includegraphics[width=1\textwidth]{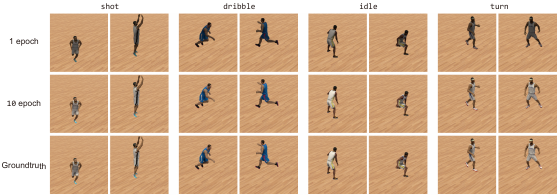}
		\caption{Qualitative comparisons on single-person scenes from the GalaBasketball dataset.} 
		\label{fig5-4}
	\end{figure}

\begin{figure}[htbp]
		\centering
		\includegraphics[width=0.90\textwidth]{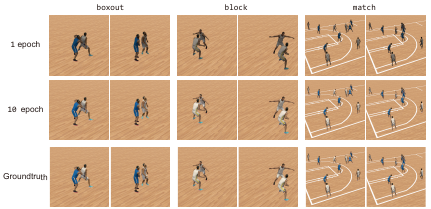}
		\caption{Qualitative comparisons on multi-person scenes from the GalaBasketball dataset.} 
		\label{fig5-5}
	\end{figure}

Figures \ref{fig5-4} and \ref{fig5-5} present qualitative visual comparisons for single-person complex motion scenarios and multi-person interactive scenes, respectively.
In single-person scenarios, benefiting from the pose-guided precise deformation and geometry-aware hash encoding, our method not only preserves a complete and smooth human surface topology but also faithfully reconstructs the dynamic shading and high-frequency clothing wrinkles during extreme motions. 
Visually, it is completely free from noticeable artifacts or over-stretching distortions.
In multi-person scenes, frequent mutual occlusions and complex depth ambiguities inherently exist among different individuals.
The visualizations vividly demonstrate that our proposed reconstruction framework can effectively and independently model multiple human instances with remarkable precision using 3D Gaussians.
Guided by depth rendering and normal constraints, our approach maintains crisp boundaries in overlapping regions, effectively preventing erroneous blending and color bleeding across the Gaussian point clouds of distinct individuals.
These compelling results strongly indicate that our 3D Gaussian reconstruction framework, grounded in region-aware strategies and geometric priors, not only excels in single-person rendering but also exhibits profound scalability and robustness when extended to complex, dynamic multi-person environments.

\subsubsection{Ablation Studies}

To validate the efficacy of the key components within our proposed framework, we conduct comprehensive ablation studies on the PeopleSnapshot dataset. The analysis spans two dimensions: the quantitative performance degradation of the network architecture and the qualitative visual discrepancies arising from different initialization templates.

Table \ref{tab5-3} reports the reconstruction performance comparisons when isolating various components of the hash encoding module. Our full method, incorporating all components, consistently achieves the optimal reconstruction metrics.
When the front and back surface normal map constraints are removed (w/o $N_f, N_b$), the model suffers a precipitous drop in PSNR, plummeting by 4.96 dB.
Similarly, omitting the multi-view depth information (w/o $D_{\{f,b,l,r\}}$) leads to a severe PSNR decline to 24.09 dB.
This highlights the critical role of incorporating global depth and normal features as auxiliary conditions during the hash feature querying. These geometric priors effectively endow the MLP with explicit spatial constraints and torso collision awareness.
Ablating the vertex displacement parameter field (w/o hash-vd) severely degrades the perceptual quality, escalating the LPIPS to 0.1523. This demonstrates that directly optimizing unconstrained Gaussian centers inevitably leads to optimization divergence and structural distortions during dynamic deformations.
Furthermore, disabling the spherical harmonics hash (w/o hash-SH) compromises color fidelity, yielding a PSNR drop to 26.53 dB. This confirms that querying appearance attributes from a continuous parameter field effectively preserves the color consistency of spatially adjacent Gaussians.
Finally, omitting the multi-scale feature query mechanism (w/o multiscale) causes the LPIPS error to surge to 0.0563.
This fundamentally indicates that a single-resolution hash grid fails to concurrently accommodate the smooth topology of the torso and the high-frequency geometric details of the facial and hand regions.

	\begin{table}[htbp]
	\centering
	\caption{Ablation Study of the Hash Encoding Module on PeopleSnapshot.}
	\resizebox{0.50\textwidth}{!}{
		\begin{tabular}{lccc}
			\toprule[1pt]
			& PSNR$\uparrow$(dB) & SSIM$\uparrow$ & LPIPS$\downarrow$ \\ \midrule[0.75pt]
			w/o hash-SH           & 26.53          & 0.9322         & 0.0720             \\
			w/o hash-vd           & 25.40           & 0.9075         & 0.1523            \\
			w/o $D_{\{f,b,l,r\}}$ & 24.09          & 0.9365         & 0.0594            \\
			w/o $N_f, N_b$        & 23.18          & 0.9248         & 0.0771            \\
			w/o multiscale        & 25.46          & 0.9462         & 0.0563            \\
			Ours                  & \textbf{28.14}          & \textbf{0.9653}         & \textbf{0.0291}            \\ \bottomrule[1pt]
		\end{tabular}
	}
	\label{tab5-3}
	\end{table}

\begin{figure}[htbp]
	\centering
	\includegraphics[width=0.85\textwidth]{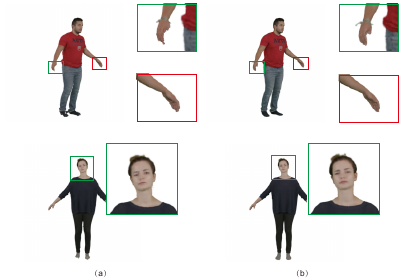}
	\caption{Detailed comparisons of facial and hand regions using different parametric templates. (a) SMPL-based initialization. (b) SMPL-X-based initialization.} 
	\label{fig5-7}
	\end{figure}

To underscore the significance of our proposed region-aware density initialization strategy predicated on the SMPL-X model, we present qualitative comparisons of the reconstructed high-frequency details in the hand and facial regions using distinct parametric templates, as visualized in Figure \ref{fig5-7}.
Due to the absence of fine-grained hand joint articulations in the SMPL model, surface sampling yields insufficient Gaussian point density in the extremities, inherently lacking pose guidance. Consequently, the rendered fingers exhibit severe fused, distorted, or blurred artifacts.
Conversely, by adopting the SMPL-X-based initialization strategy—which performs dense sampling within the local neighborhood of hand vertices and seamlessly incorporates hand joint rotation parameters—the reconstructed finger contours emerge distinctly crisper and sharper.
Furthermore, the intrinsic inability of the SMPL model to represent facial expressions inevitably leads to a markedly blurred texture during facial reconstruction.
Empowered by the expressive parameters and localized high-density sampling of SMPL-X, our method not only faithfully restores the intricate facial topology of the subject but also precisely captures subtle undulations around the eyes, nose bridge, and overall facial structure, thereby profoundly elevating the visual fidelity of the synthesized human avatars.

\section{Conclusion}

	In this paper, we introduced a novel 3D Gaussian human reconstruction framework that integrates the robust geometric priors of the SMPL-X model with the efficient representations of continuous spatial fields.
	It utilizes a region-aware formulation to enhance high-fidelity 3D human avatar recovery from monocular RGB images, particularly in dynamic scenarios featuring complex non-rigid deformations.
	Specifically, the SMPL-X guided initialization acts to establish a generic and robust geometric foundation for 3D Gaussians and skinning weights, while the region-aware density initialization functions to intrinsically preserve high-frequency details in critical areas such as the face and hands.
	To effectively balance rendering fidelity and memory footprint, the geometry-aware multi-scale hash encoding module aligns spatial representations with dynamic attributes, and the decoupled parameter field facilitates the photorealistic rendering of anisotropic Gaussian distributions across diverse poses.
	Experiments on standard benchmark datasets affirm the efficacy and superior detail recovery of our proposed model compared to state-of-the-art techniques in complex real-world motions, while strictly maintaining rapid rendering speeds.

	\textbf{Limitations and future work.} 
	While our proposed framework achieves high-fidelity dynamic 3D human reconstruction and successfully mitigates localized artifacts, it still presents certain limitations that warrant further investigation. 
	First, our method inherently relies on the SMPL-X model to provide geometric priors and skinning weights. Consequently, the reconstruction quality is susceptible to the accuracy of the initial pose and shape estimation. 
	In unconstrained scenarios with extreme poses or severe self-occlusions, erroneous SMPL-X parameter estimation may introduce initial geometric deviations that are challenging to fully correct during the subsequent optimization process. 
	Second, since the skinning weights are derived from a parametric body model, representing highly complex topological variations—such as loose, flowing garments, long dynamic hair, or hand-object interactions—remains a persistent challenge. 
	Furthermore, due to the ill-posed nature of monocular reconstruction, accurately recovering high-frequency textures for persistently unseen or heavily occluded regions can still lead to slight over-smoothing, despite our region-aware compensation.

	To address these limitations, our future work will focus on several promising directions. 
	Primarily, we plan to introduce a joint optimization mechanism that dynamically refines the underlying SMPL-X parameters concurrently with the 3D Gaussian attributes, thereby enhancing the system's robustness against imperfect initializations. 
	Additionally, incorporating physics-based priors or garment-specific dynamic templates could significantly improve the modeling of loose clothing and complex topologies. 
	Finally, exploring the integration of powerful generative foundation models (e.g., conditional diffusion models) to plausibly hallucinate geometric and textural details for unseen regions, along with developing Gaussian compression techniques for efficient deployment on edge devices, constitute exciting avenues for future research.




\small
\bibliographystyle{elsarticle-num}
\bibliography{refs.bib}

@inproceedings{cao2022bilateral,
	title={Bilateral normal integration},
	author={Cao, Xu and Santo, Hiroaki and Shi, Boxin and Okura, Fumio and Matsushita, Yasuyuki},
	booktitle={Proceedings of the European Conference on Computer Vision (ECCV)},
	pages={552--567},
	year={2022},
	organization={Springer}
}

@article{kerbl20233d,
	title={3d gaussian splatting for real-time radiance field rendering},
	author={Kerbl, Bernhard and Kopanas, Georgios and Leimk\"uhler, Thomas and Drettakis, George and others},
	journal={ACM Transactions on Graphics (TOG)},
	volume={42},
	number={4},
	pages={139:1--139:14},
	year={2023}
}

@inproceedings{liu2024animatable,
	title={Animatable 3d gaussian: Fast and high-quality reconstruction of multiple human avatars},
	author={Liu, Yang and Huang, Xiang and Qin, Minghan and Lin, Qinwei and Wang, Haoqian},
	booktitle={Proceedings of the 32nd ACM International Conference on Multimedia},
	pages={1120--1129},
	year={2024}
}

@article{muller2022instant,
	title={Instant neural graphics primitives with a multiresolution hash encoding},
	author={M\"uller, Thomas and Evans, Alex and Schied, Christoph and Keller, Alexander},
	journal={ACM Transactions on Graphics (TOG)},
	volume={41},
	number={4},
	pages={1--15},
	year={2022},
	publisher={ACM New York, NY, USA}
}

@article{mildenhall2021nerf,
	title={Nerf: Representing scenes as neural radiance fields for view synthesis},
	author={Mildenhall, Ben and Srinivasan, Pratul P and Tancik, Matthew and Barron, Jonathan T and Ramamoorthi, Ravi and Ng, Ren},
	journal={Communications of the ACM},
	volume={65},
	number={1},
	pages={99--106},
	year={2021},
	publisher={ACM New York, NY, USA}
}

@inproceedings{alldieck2018detailed,
	title={Detailed human avatars from monocular video},
	author={Alldieck, Thiemo and Magnor, Marcus and Xu, Weipeng and Theobalt, Christian and Pons-Moll, Gerard},
	booktitle={2018 International Conference on 3D Vision (3DV)},
	pages={98--109},
	year={2018},
	organization={IEEE}
}

@article{wang2004image,
	title={Image quality assessment: from error visibility to structural similarity},
	author={Wang, Zhou and Bovik, Alan C and Sheikh, Hamid R and Simoncelli, Eero P},
	journal={IEEE Transactions on Image Processing (TIP)},
	volume={13},
	number={4},
	pages={600--612},
	year={2004},
	publisher={IEEE}
}

@inproceedings{jiang2023instantavatar,
	title={Instantavatar: Learning avatars from monocular video in 60 seconds},
	author={Jiang, Tianjian and Chen, Xu and Song, Jie and Hilliges, Otmar},
	booktitle={Proceedings of the IEEE/CVF Conference on Computer Vision and Pattern Recognition (CVPR)},
	pages={16922--16932},
	year={2023}
}

@article{chen2021animatable,
	title={Animatable neural radiance fields from monocular rgb videos},
	author={Chen, Jianchuan and Zhang, Ying and Kang, Di and Zhe, Xuefei and Bao, Linchao and Jia, Xu and Lu, Huchuan},
	journal={arXiv preprint arXiv:2106.13629},
	year={2021}
}

@inproceedings{peng2021neural,
	title={Neural body: Implicit neural representations with structured latent codes for novel view synthesis of dynamic humans},
	author={Peng, Sida and Zhang, Yuanqing and Xu, Yinghao and Wang, Qianqian and Shuai, Qing and Bao, Hujun and Zhou, Xiaowei},
	booktitle={Proceedings of the IEEE/CVF Conference on Computer Vision and Pattern Recognition (CVPR)},
	pages={9054--9063},
	year={2021}
}

@inproceedings{xiu2022icon,
	title={Icon: Implicit clothed humans obtained from normals},
	author={Xiu, Yuliang and Yang, Jinlong and Tzionas, Dimitrios and Black, Michael J},
	booktitle={Proceedings of the IEEE/CVF Conference on Computer Vision and Pattern Recognition (CVPR)},
	pages={13286--13296},
	year={2022},
	organization={IEEE}
}

@inproceedings{xiu2023econ,
	title={Econ: Explicit clothed humans optimized via normal integration},
	author={Xiu, Yuliang and Yang, Jinlong and Cao, Xu and Tzionas, Dimitrios and Black, Michael J},
	booktitle={Proceedings of the IEEE/CVF Conference on Computer Vision and Pattern Recognition (CVPR)},
	pages={512--523},
	year={2023}
}

@inproceedings{fridovich2022plenoxels,
	title={Plenoxels: Radiance fields without neural networks},
	author={Fridovich-Keil, Sara and Yu, Alex and Tancik, Matthew and Chen, Qinhong and Recht, Benjamin and Kanazawa, Angjoo},
	booktitle={Proceedings of the IEEE/CVF Conference on Computer Vision and Pattern Recognition (CVPR)},
	pages={5501--5510},
	year={2022}
}

@inproceedings{sun2022direct,
	title={Direct voxel grid optimization: Super-fast convergence for radiance fields reconstruction},
	author={Sun, Cheng and Sun, Min and Chen, Hwann-Tzong},
	booktitle={Proceedings of the IEEE/CVF Conference on Computer Vision and Pattern Recognition (CVPR)},
	pages={5459--5469},
	year={2022}
}

@inproceedings{wang2022fourier,
	title={Fourier plenoctrees for dynamic radiance field rendering in real-time},
	author={Wang, Liao and Zhang, Jiakai and Liu, Xinhang and Zhao, Fuqiang and Zhang, Yanshun and Zhang, Yingliang and Wu, Minye and Yu, Jingyi and Xu, Lan},
	booktitle={Proceedings of the IEEE/CVF Conference on Computer Vision and Pattern Recognition (CVPR)},
	pages={13524--13534},
	year={2022}
}

@inproceedings{zwicker2001ewa,
	title={Ewa volume splatting},
	author={Zwicker, Matthias and Pfister, Hanspeter and {Van} Baar, Jeroen and Gross, Markus},
	booktitle={Proceedings Visualization, 2001. VIS'01.},
	pages={29--538},
	year={2001},
	organization={IEEE}
}

@inproceedings{zhang2018unreasonable,
	title={The unreasonable effectiveness of deep features as a perceptual metric},
	author={Zhang, Richard and Isola, Phillip and Efros, Alexei A and Shechtman, Eli and Wang, Oliver},
	booktitle={Proceedings of the IEEE/CVF Conference on Computer Vision and Pattern Recognition (CVPR)},
	pages={586--595},
	year={2018}
}

@incollection{loper2023smpl,
	title={SMPL: A skinned multi-person linear model},
	author={Loper, Matthew and Mahmood, Naureen and Romero, Javier and Pons-Moll, Gerard and Black, Michael J},
	booktitle={Seminal Graphics Papers: Pushing the Boundaries, Volume 2},
	pages={851--866},
	year={2023}
}

@inproceedings{pavlakos2019expressive,
	title={Expressive body capture: 3d hands, face, and body from a single image},
	author={Pavlakos, Georgios and Choutas, Vasileios and Ghorbani, Nima and Bolkart, Timo and Osman, Ahmed AA and Tzionas, Dimitrios and Black, Michael J},
	booktitle={Proceedings of the IEEE/CVF Conference on Computer Vision and Pattern Recognition (CVPR)},
	pages={10975--10985},
	year={2019}
}

@inproceedings{bai2023learning,
	title={Learning personalized high quality volumetric head avatars from monocular rgb videos},
	author={Bai, Ziqian and Tan, Feitong and Huang, Zeng and Sarkar, Kripasindhu and Tang, Danhang and Qiu, Di and Meka, Abhimitra and Du, Ruofei and Dou, Mingsong and Orts-Escolano, Sergio and others},
	booktitle={Proceedings of the IEEE/CVF Conference on Computer Vision and Pattern Recognition},
	pages={16890--16900},
	year={2023}
}

@article{bharadwaj2023flare,
	title={Flare: Fast learning of animatable and relightable mesh avatars},
	author={Bharadwaj, Shrisha and Zheng, Yufeng and Hilliges, Otmar and Black, Michael J and Fernandez-Abrevaya, Victoria},
	journal={arXiv preprint arXiv:2310.17519},
	year={2023}
}

@inproceedings{chai2022realy,
	title={Realy: Rethinking the evaluation of 3d face reconstruction},
	author={Chai, Zenghao and Zhang, Haoxian and Ren, Jing and Kang, Di and Xu, Zhengzhuo and Zhe, Xuefei and Yuan, Chun and Bao, Linchao},
	booktitle={European conference on computer vision},
	pages={74--92},
	year={2022},
	organization={Springer}
}

@inproceedings{chen2021snarf,
	title={Snarf: Differentiable forward skinning for animating non-rigid neural implicit shapes},
	author={Chen, Xu and Zheng, Yufeng and Black, Michael J and Hilliges, Otmar and Geiger, Andreas},
	booktitle={Proceedings of the IEEE/CVF International Conference on Computer Vision},
	pages={11594--11604},
	year={2021}
}

@article{chen2023fast,
	title={Fast-snarf: A fast deformer for articulated neural fields},
	author={Chen, Xu and Jiang, Tianjian and Song, Jie and Rietmann, Max and Geiger, Andreas and Black, Michael J and Hilliges, Otmar},
	journal={IEEE Transactions on Pattern Analysis and Machine Intelligence},
	volume={45},
	number={10},
	pages={11796--11809},
	year={2023},
	publisher={IEEE}
}

@article{collet2015high,
	title={High-quality streamable free-viewpoint video},
	author={Collet, Alvaro and Chuang, Ming and Sweeney, Pat and Gillett, Don and Evseev, Dennis and Calabrese, David and Hoppe, Hugues and Kirk, Adam and Sullivan, Steve},
	journal={ACM Transactions on Graphics (ToG)},
	volume={34},
	number={4},
	pages={1--13},
	year={2015},
	publisher={ACM New York, NY, USA}
}

@inproceedings{grassal2022neural,
	title={Neural head avatars from monocular rgb videos},
	author={Grassal, Philip-William and Prinzler, Malte and Leistner, Titus and Rother, Carsten and Nie{\ss}ner, Matthias and Thies, Justus},
	booktitle={Proceedings of the IEEE/CVF conference on computer vision and pattern recognition},
	pages={18653--18664},
	year={2022}
}

@inproceedings{guo2023vid2avatar,
	title={Vid2avatar: 3d avatar reconstruction from videos in the wild via self-supervised scene decomposition},
	author={Guo, Chen and Jiang, Tianjian and Chen, Xu and Song, Jie and Hilliges, Otmar},
	booktitle={Proceedings of the IEEE/CVF Conference on Computer Vision and Pattern Recognition},
	pages={12858--12868},
	year={2023}
}

@article{habermann2023hdhumans,
	title={Hdhumans: A hybrid approach for high-fidelity digital humans},
	author={Habermann, Marc and Liu, Lingjie and Xu, Weipeng and Pons-Moll, Gerard and Zollhoefer, Michael and Theobalt, Christian},
	journal={Proceedings of the ACM on Computer Graphics and Interactive Techniques},
	volume={6},
	number={3},
	pages={1--23},
	year={2023},
	publisher={ACM New York, NY, USA}
}

@inproceedings{ho2023learning,
	title={Learning locally editable virtual humans},
	author={Ho, Hsuan-I and Xue, Lixin and Song, Jie and Hilliges, Otmar},
	booktitle={Proceedings of the IEEE/cvf conference on computer vision and pattern recognition},
	pages={21024--21035},
	year={2023}
}

@article{li2017learning,
	title={Learning a model of facial shape and expression from 4D scans.},
	author={Li, Tianye and Bolkart, Timo and Black, Michael J and Li, Hao and Romero, Javier},
	journal={ACM Trans. Graph.},
	volume={36},
	number={6},
	pages={194--1},
	year={2017}
}

@inproceedings{ma2020learning,
	title={Learning to dress 3d people in generative clothing},
	author={Ma, Qianli and Yang, Jinlong and Ranjan, Anurag and Pujades, Sergi and Pons-Moll, Gerard and Tang, Siyu and Black, Michael J},
	booktitle={Proceedings of the IEEE/CVF Conference on Computer Vision and Pattern Recognition},
	pages={6469--6478},
	year={2020}
}

@inproceedings{ma2021pixel,
	title={Pixel codec avatars},
	author={Ma, Shugao and Simon, Tomas and Saragih, Jason and Wang, Dawei and Li, Yuecheng and De La Torre, Fernando and Sheikh, Yaser},
	booktitle={Proceedings of the IEEE/CVF Conference on Computer Vision and Pattern Recognition},
	pages={64--73},
	year={2021}
}

@inproceedings{peng2021animatable,
	title={Animatable neural radiance fields for modeling dynamic human bodies},
	author={Peng, Sida and Dong, Junting and Wang, Qianqian and Zhang, Shangzhan and Shuai, Qing and Zhou, Xiaowei and Bao, Hujun},
	booktitle={Proceedings of the IEEE/CVF international conference on computer vision},
	pages={14314--14323},
	year={2021}
}

@inproceedings{prokudin2021smplpix,
	title={Smplpix: Neural avatars from 3d human models},
	author={Prokudin, Sergey and Black, Michael J and Romero, Javier},
	booktitle={Proceedings of the IEEE/CVF winter conference on applications of computer vision},
	pages={1810--1819},
	year={2021}
}

@inproceedings{shen2023x,
	title={X-avatar: Expressive human avatars},
	author={Shen, Kaiyue and Guo, Chen and Kaufmann, Manuel and Zarate, Juan Jose and Valentin, Julien and Song, Jie and Hilliges, Otmar},
	booktitle={Proceedings of the IEEE/CVF Conference on Computer Vision and Pattern Recognition},
	pages={16911--16921},
	year={2023}
}

@article{xiang2021modeling,
	title={Modeling clothing as a separate layer for an animatable human avatar},
	author={Xiang, Donglai and Prada, Fabian and Bagautdinov, Timur and Xu, Weipeng and Dong, Yuan and Wen, He and Hodgins, Jessica and Wu, Chenglei},
	journal={ACM Transactions on Graphics (TOG)},
	volume={40},
	number={6},
	pages={1--15},
	year={2021},
	publisher={ACM New York, NY, USA}
}

@article{xu2021h,
	title={H-nerf: Neural radiance fields for rendering and temporal reconstruction of humans in motion},
	author={Xu, Hongyi and Alldieck, Thiemo and Sminchisescu, Cristian},
	journal={Advances in Neural Information Processing Systems},
	volume={34},
	pages={14955--14966},
	year={2021}
}

@inproceedings{yu2023monohuman,
	title={Monohuman: Animatable human neural field from monocular video},
	author={Yu, Zhengming and Cheng, Wei and Liu, Xian and Wu, Wayne and Lin, Kwan-Yee},
	booktitle={Proceedings of the IEEE/CVF Conference on Computer Vision and Pattern Recognition},
	pages={16943--16953},
	year={2023}
}

@inproceedings{zielonka2023instant,
	title={Instant volumetric head avatars},
	author={Zielonka, Wojciech and Bolkart, Timo and Thies, Justus},
	booktitle={Proceedings of the IEEE/CVF conference on computer vision and pattern recognition},
	pages={4574--4584},
	year={2023}
}

@article{xie2025nsghg,
	title={NSGHG: Neural surface guided generalizable human Gaussian splatting for sparse view synthesis},
	author={Xie, Hong and Zhang, Xiaoyan and Zhu, Yingying},
	journal={Neurocomputing},
	volume={653},
	pages={131207},
	year={2025},
	publisher={Elsevier}
}

@article{nazir20263dgeomeshnet,
	title={3DGeoMeshNet: A multi-scale graph auto-encoder for 3D mesh reconstruction and completion},
	author={Nazir, Saqib and L{\'e}zoray, Olivier and Bougleux, S{\'e}bastien},
	journal={Neurocomputing},
	pages={132652},
	year={2026},
	publisher={Elsevier}
}

@article{li2026novel,
	title={Novel view synthesis for underwater scene with Gaussian splat fields and physically-based water modeling},
	author={Li, Qian and Fu, Rao},
	journal={Neurocomputing},
	pages={133060},
	year={2026},
	publisher={Elsevier}
}

@article{wang2026dynamic,
	title={Dynamic view synthesis with topologically-varying neural radiance fields from sparse input views},
	author={Wang, Kangkan and Wei, Kejie and Li, Shao-Yuan},
	journal={Neurocomputing},
	pages={132942},
	year={2026},
	publisher={Elsevier}
}





\end{document}